\documentclass{jpsj-suppl}
\usepackage{txfonts} 
\usepackage{graphicx}
\usepackage{epsfig}

\newlength{\figwidth}
\newlength{\shift}
\newcommand{\fg}[3]
{
\begin{figure}[ht]

\vspace*{-0cm}
\[
\includegraphics[width=\figwidth]{#1}
\]
\vskip -0.2cm
\caption{\label{#2}
\small#3
}
\end{figure}}

\newcommand{\dg}{^{\dagger }}

\newcommand{\up}{\uparrow}
\newcommand{\down}{\downarrow}

\newcommand \bea {\begin{eqnarray} }
\newcommand \eea {\end{eqnarray}}
\newcommand{\bk}{{\bf{k}}}

\newcommand{\bQ}{{\bf{Q}}}
\newcommand{\bR}{{\bf{R}}}
\newcommand{\urs}{URu$_{2}$Si$_{2}$\ }
\newcommand{\ursp}{URu$_{2}$Si$_{2}$}

\newcommand{\psic}{{c}}

\title{Hidden and Hastatic Orders in URu$_2$Si$_2$}

\author{Rebecca \textsc{Flint}$^{1,2}$, Premala \textsc{Chandra}$^{3}$ and Piers \textsc{Coleman}$^{3,4}$}

\inst{$^{1}$Department of Physics, Massachusetts Institute of Technology, 77 Massachusetts Avenue, Cambridge, Massachusetts 02139 USA\\
$^{2}$Department of Physics and Astronomy, Iowa State University, 12 Physics Hall, Ames, Iowa 50011 USA\\
$^{3}$Center for Materials Theory, Department of Physics and Astronomy, Rutgers University, 136 Frelinghuysen Rd, Piscataway, New Jersey 08854 USA\\
$^{4}$Department of Physics, Royal Holloway, University of London, Egham, Surrey, TW20 0EX, UK}

\email{flint@iastate.edu}

\recdate{October 20, 2013}

\abst{The hidden order developing below 17.5K in the heavy fermion material URu$_2$Si$_2$ has eluded identification for over twenty five years.  This paper will review the recent theory of ``hastatic order,'' a novel two-component order parameter capturing the hybridization between half-integer spin (Kramers) conduction electrons and the non-Kramers 5f$^2$ Ising local moments, as strongly indicated by the observation of Ising quasiparticles in de Haas-van Alphen measurements.  Hastatic order differs from conventional magnetism as it is a spinor order that breaks both single and double time-reversal symmetry by mixing states of different Kramers parity.  The broken time-reversal symmetry simply explains both the pseudo-Goldstone mode between the hidden order and antiferromagnetic phases and the nematic order seen in torque magnetometry.  The spinorial nature of the hybridization also explains how the Kondo effect gives a phase transition, with the hybridization gap turning on at the hidden order transition as seen in scanning tunneling microscopy.  Hastatic order also has a number of new predictions: a basal-plane magnetic moment of order .01$\mu_B$, a gap to longitudinal spin fluctuations that vanishes continuously at the first order antiferromagnetic transition and a narrow resonant nematic feature in the scanning tunneling spectra.}

\kword{URu$_2$Si$_2$, hidden order, Kondo effect}

\begin{document}
\maketitle

\section{Introduction}

The nature of the ``hidden order'' in \urs is one of the oldest open problems in strongly correlated electrons\cite{Palstra85,Schlabitz86,Mydosh11}.  There is a clear mean-field-like specific heat jump at $T_{HO} = 17.5$K, which, according to the entropy $S(T_{HO}) \sim 1/3 R \log 2$ should correspond to a large order parameter.  However, despite years of active research, the nature of this order parameter is still unknown, suggesting that some fundamentally new phenomenon is responsible.  Recently, we proposed hastatic order as this phenomenon\cite{Chandra13}.  Hastatic order is a type of spinorial hybridization due to the hybridization of a non-Kramers (integer spin) doublet with Kramers (half-integer) conduction electrons.  Much of the complex phenomenology of \urs emerges naturally out of this simple idea.  For example, although the hybridization gap is relatively large, the magnitude of any observable moments is limited by the Kondo effect.  We can also explain the Ising anisotropy of the heavy quasiparticles deep within the hidden order; the broken tetragonal symmetry seen both in the spin and charge channels;  and the pseudo-Goldstone mode seen in inelastic neutron scattering.  This article is intended mainly as a review article for non-experts; for details, please refer to our previous paper\cite{Chandra13}.  

\section{Background}

\subsection{Time-reversal Symmetry}

While classically, spins simply invert under time-reversal, $\vec{S} \xrightarrow{\Theta} -\vec{S}$, quantum mechanically time-reversal is an anti-unitary operator acting on the wave-function\cite{sakurai}.  For spin-1/2 objects or fermions, that wave-function is a spinor, where a single operation of time-reversal flips the spin and two operations of time-reversal returns the original spinor multiplied by a negative phase factors. It takes four time-reversal operations to return to the original wavefunction.  By contrast, integer spin or bosonic wavefunctions invert under a single time-reversal operation and are invariant under double time-reversal symmetry (like a vector).

In heavy fermion materials, strong spin-orbit coupling makes $J$ the relevant quantum number, and crystal fields split the $2J+1$ levels into multiplets. Here, the difference between integer and half-integer spins has an important consequence: Kramers theorem, which guarantees that half integer multiplets must be at least doubly degenerate if time-reversal is preserved.  Ions with half-integer $J$'s (due to an odd number of f-electrons) are called Kramers ions, and the resulting doublets are called Kramers doublets.  Ions with integer $J$'s (an even number of f-electrons) are called non-Kramers ions and have no such protection.  Consequentially, while there may be ``non-Kramers'' doublets, these can be split into singlets by lowering the crystal symmetry\cite{sakurai}.  Non-Kramers doublets may be completely non-magnetic (like the $\Gamma_3$ quadrupolar doublet), or they may be magnetic, but are typically Ising-like in that $\langle J_z\rangle$ is dipolar, but $\langle J_\perp\rangle$ is some higher order multipole\cite{CoxZawad}.  For example, the relevant doublet for $5f^2$ ($J = 4$) in tetragonal symmetry is the $\Gamma_5$ doublet\cite{Amitsuka94} (the foundation of hastatic order):
\begin{equation}
|\Gamma_5 \pm\rangle = \alpha |J_z = \pm 3\rangle + \beta |J_z =\mp 1\rangle,
\end{equation}
and is protected by both tetragonal and time-reversal symmetries;
when either of these is broken, it splits.  The c-axis moment is magnetic, while the in-plane moments are $xy$ and $x^2-y^2$ quadrupoles. 

\subsection{Kondo Physics and Hybridization}

Heavy fermion materials like \urs combine nearly free conduction electrons with a lattice of localized f-electrons forming magnetic moments.  These two systems are decoupled at high temperatures, giving rise to a Curie-like magnetic susceptibility.  However, as the temperature decreases, they are coupled by the Kondo effect, an antiferromagnetic interaction by which the conduction electrons eventually screen out the local moments, forming a heavy Fermi liquid. We can think of this interaction as the hybridization of a dispersive conduction band with a flat (localized) band of f-electrons (see Fig 1(b)).  These are, of course, not the original f-electrons, as those charge degrees of freedom were frozen out at much higher temperatures; instead these are ``composite fermions'', which combine the creation of a conduction electron with a spin-flip, $f\dg_\up \sim c\dg_\down S_{f+}$\cite{PiersReview}.  At low temperatures, these two bands hybridize, with a temperature dependent hybridization $V \sim \langle c\dg f\rangle$ (see  Fig. 1b).  The resulting band is much flatter, corresponding to a higher effective mass, $m^*$.  There is also a small hybridization gap, $\Delta_H$ typically either above or below the Fermi energy.  This hybridization develops as a crossover around the Kondo temperature, $T_K$, and the width of the hybridization gap, $\Delta_H$ is proportional to $T_K$.

\figwidth=15cm
\fg{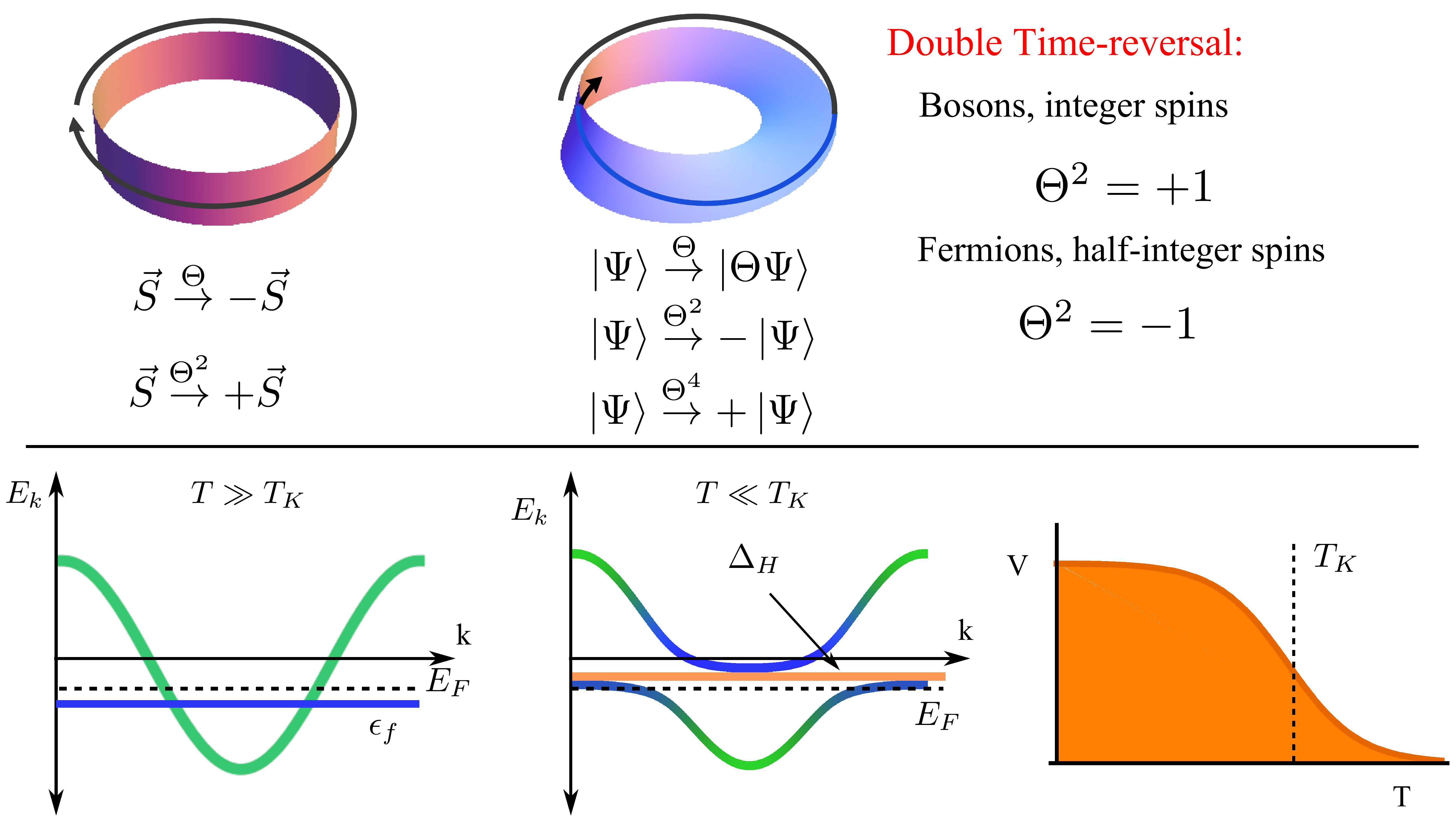}{hybrid}{(a) A spinor is like the square root of a vector: while vectors, like $\vec{S}$ reverse under a single operation of time-reversal, $\theta$, spinors like $|\hat \Psi\rangle$ reverse under two time-reversal operations.  These different behaviors characterize integer and half-integer spins, respectively.  All normal time-reversal-symmetry breaking order parameters behave like vectors under time-reversal. (b) The development of heavy fermion behavior is a hybridization between dispersive conduction electrons (green) and localized f-electrons* (blue).  The left figure shows the unhybridized bands, which will hybridize with a  hybridization, V that develops as a crossover below $T_K$ (see right figure) to form two dispersive hybridized bands.  The new Fermi level sits in the flatter part of the band, leading to a heavy band mass, and a hybridization gap, $\Delta_H$ either above or below the Fermi level (depending on whether the hybridization is with f-electrons or f-holes).  $\Delta_H$ is proportional to $T_K$,  and much smaller than the low temperature magnitude of $V = \sqrt{D T_K}$, where $D$ is the conduction electron bandwidth.}

The Kondo interaction comes from virtual valence fluctuations between a ground state doublet and excited singlet states: for a $5f^1$ state, an electron can hop off the f-ion (into the conduction sea), leaving the f-ion in an excited $5f^0$ state. Then when another electron hops back on, it can do so with the opposite spin.   Modeling these systems theoretically is challenging because spins do not obey Wick's theorem and cannot be simply treated with quantum field theoretic techniques.  To overcome this difficulty, we represent the spin with pseudo-fermions, $\vec{S}_f = \frac{1}{2}f\dg_\alpha \vec{\sigma}_{\alpha \beta} f_\beta$ and the excited $5f^0$ state with a slave boson, $|0\rangle = b\dg |\Omega\rangle$\cite{Coleman83}.  In the mean-field theory, the slave bosons condense below $T_K$ and the hybridization is proportional to $\langle b \rangle$.  This condensation incorrectly gives a phase transition at $T_K$, but as no symmetries are broken, going beyond mean field theory restores it to a crossover\cite{Read83}. 

\subsection{Hidden Order Background and Relevant Experiments}

\urs is a heavy fermion material with Ising spins on the U sites.  Around 70K, these moments  begin to be screened by the conduction electrons, indicating the partial formation of a heavy Fermi liquid.  At $T_{HO} = 17.5$K, there is a second order phase transition into the hidden ordered (HO) state, which gaps out large portions of the Fermi surface; the large entropy involved in this transition indicates that the local moments have only been partially screened by $T_{HO}$\cite{Palstra85}.   Under pressure, the hidden order undergoes a first order transition to an antiferromagnetic (AFM)  phase with an ordering vector $\bQ = [001]$\cite{Jo07,Villaume08,Hassinger10}.  de Haas-van Alphen experiments find the same Fermi surface in both the AFM and HO states, implying the two states have the same $\bQ$\cite{Hassinger10}.  Inelastic neutron scattering has shown a softening longitidunal mode interpreted as a pseudo-Goldstone mode between the HO and AFM\cite{Broholm91,Niklowitz11}.  This interpretation has the startling implication that HO likely also breaks time-reversal symmetry, as there are not typically Goldstone modes between two phases with different time-reversal properties.  Furthermore, broken tetragonal symmetry developing at $T_{HO}$ has been seen in torque magnetometry data (as a $\chi_{xy}(T)$) and as a small orthorhombic distortion\cite{Okazaki11,Matsuda}.

There are two sets of recent experiments that together suggest hastatic order in \ursp.  The first are recent scanning tunneling microscopy (STM) experiments that use quasiparticle interference to measure the bandstructure and can see the development of hybridization between conduction electrons and local moments as a band-bending that develops precisely at $T_{HO}$\cite{Schmidt10, Aynajian10}.  The hybridization gap measured in STM has the same magnitude as the specific heat gap, and so the HO gap can be explained as a hybridization gap that partially gaps out the Fermi surface, and develops at $T_{HO}$ as an order parameter.  ARPES experiments also indicate the development of coherent quasiparticles at $T_{HO}$\cite{lanl,Shen13}, and together these suggest that HO is the development of hybridization.  Of course, normally hybridization only develops as a crossover, not a phase transition, and so there must be something quite unusual (and symmetry breaking) about this hybridization.

The second set of experiments are two measurements of the magnetic anisotropy of the heavy quasiparticles in the HO state.  de Haas-van Alphen measurements in a tilted magnetic field show that the g-factor of the quasiparticles is distinctly Ising-like with a $\cos \theta$ dependence\cite{Ohkuni99, Altarawneh11}.  The superconducting $H_{C2}$ has a similar Ising anisotropy\cite{Hc2, Altarawneh2}.  As non-interacting conduction electrons will be only weakly anisotropic, these quasiparticles must have inherited this Ising nature from the U moments; this hybridization can only happen if the U ground state is either a doublet or a ``pseudo-doublet'' formed from two singlets whose splitting is much smaller than $T_{HO}$

\section{Hybridizing with a Non-Kramers Doublet}

So we have seen that the hidden order seems to involve the hybridization of conduction electrons with an Ising local moment.  The valence of the U ion is controversial, with inelastic neutron measurements\cite{Broholm91,Park02}  favoring 5f$^2$ and EELs favoring 5f$^3$\cite{Jeffries10}.   In a tetragonal system like \ursp, the Ising nature is most consistent with a non-Kramers doublet (an Ising-like Kramers doublet is possible, but requires extreme fine-tuning\cite{Flint13}; similarly two closely spaced singlets also require fine-tuning).   So we conclude that the conduction electrons are hybridizing with a non-Kramers doublet, which has serious consequences for the nature of the hybridization.

Before we get to the microscopic picture, let us examine the simplest possible Hamiltonian mixing Kramers and non-Kramers states.  For conciseness, we represent $\theta^2$ as a $2\pi$ rotation in $SU(2)$ parameter space.  As conduction electrons are Kramers-like, their wavefunction picks up a negative sign under double time-reversal, $|\bk \sigma\rangle^{2\pi} = -|\bk \sigma \rangle$, while a non-Kramers state will not, $|Jm\rangle^{2\pi} = + |Jm\rangle$.  The mixing between these two states is given by a hybridization, $V_{m\sigma}$:
\bea
H & = & V_{m\sigma}|J m\rangle \langle \bk \sigma| + \mathrm{H.c.}\cr
& = & V_{m\sigma}^{2\pi}|J m\rangle^{2\pi}\langle \bk \sigma|^{2\pi} + \mathrm{H.c.}\cr
& = & -V_{m\sigma}^{2\pi}|J m\rangle \langle \bk \sigma| + \mathrm{H.c.}
\eea
Since the Hamiltonian must be invariant under double time-reversal, $V_{m\sigma}^{2\pi} = - V_{m\sigma}$ must pick up a negative sign under time-reversal symmetry, just like a spinor.  In fact, this spinorial nature of the hybridization should not be surprising: we are mixing a half-integer spin state with an integer spin state, and therefore the hybridization itself \emph{must} carry a half-integral angular momentum.  In a real material, this kind of mixing occurs via valence fluctuations, and as the non-Kramers state must involve an even number of f-electrons, the mixing term $V$ will involve the destruction of a Kramers state and the hybridization necessarily inherits the spinorial nature of the Kramers state; by contrast, the usual hybridization destroys a non-Kramers state and will likely not break time-reversal.

\figwidth=12cm
\fg{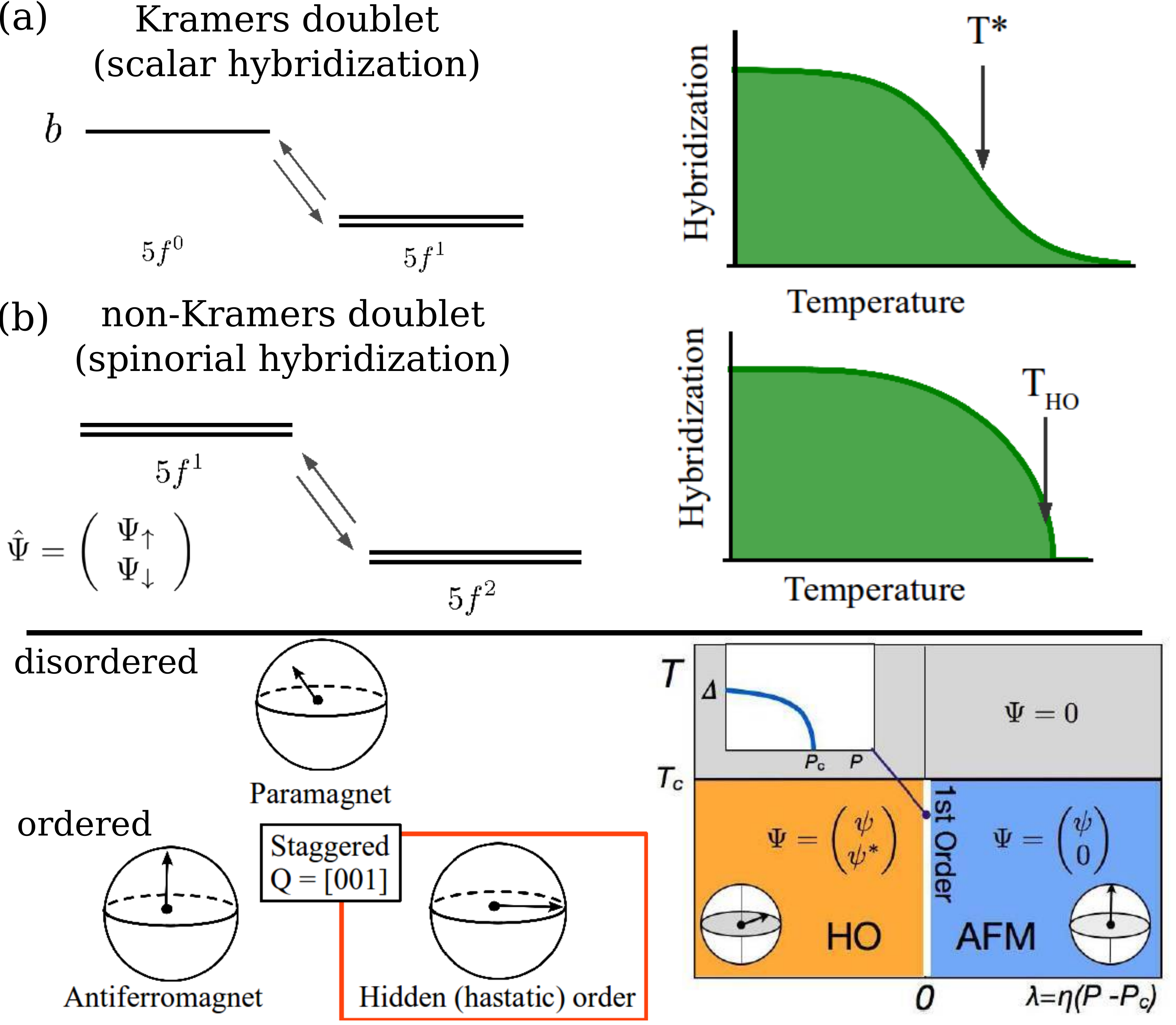}{spinorial}{(a) Normal (scalar) versus spinorial hybridization.  The hybridization of a Kramers doublet with conduction electrons will generically take place via valence fluctuations to an excited singlet, which can be represented by a single slave boson.  Therefore, the normal Kondo effect breaks no symmetries and will develop as a crossover.  By contract, the hybridization of a non-Kramers doublet with conduction electrons will \emph{necessarily} involve valence fluctuations to an excited Kramers doublet, which must be represented by a spinor of slave bosons.  This spinorial hybridization breaks time-reversal symmetry (as it must pick a direction in the excited Kramers doublet) and therefore develops as a phase transition.  (b) Once the hybridization spinor acquires a magnitude, it can either behave as a disordered paramagnet; point along the c-axis, which leads to a phase that behaves like an Ising antiferromagnet (assuming a staggered orientation); or point in the basal plane, which leads to a phase that looks suspiciously like hidden order.  This third option is what we call hastatic order. Simply by writing down a Landau-Ginzburg theory associated with the spinorial order parameter, $\Psi$, we can reproduce the basic phase diagram of  URu$_2$Si$_2$, as we can introduce a term $-\lambda \Psi^\dagger \sigma^z \Psi$ to tune between the antiferromagnet, with $\Psi$ aligned along the c-axis and hastatic order, with $\Psi$ pointing in the basal plane.  A unique consequence of this spinorial order is that the gap to longitudinal spin fluctuations vanishes as $\Delta \sim \sqrt{P-P_c}$, even though $P_C$ describes a first order phase transition.}

Turning now to the microscopic picture [shown in Fig. 2], we come to the same conclusion.  In the normal Kondo case, we represent the excited singlet with a single slave boson.  But when the ground state is a non-Kramers ion ($n$ even), all valence fluctuations will be to a Kramers ion ($n$ odd).  These excited states are protected by time-reversal symmetry and therefore must be Kramers doublets; now we have two Kondo channels instead of one. Instead of representing a single state by a slave boson $b$, we need a slave boson for each state, $\Psi_\up$ and $\Psi_\down$.  These form a spinor, and when the bosons condense (indicating the development of hybridization), this spinor describes a direction, thereby breaking time-reversal symmetry.  This spinor of slave bosons is quite similar to the Schwinger bosons used in frustrated magnetic systems, where the condensation of the bosons indicates magnetic order.  Here, the amplitude of the spinor squared is not fixed to $2S$, but instead will be given by the deviation of the valence from $n = 2$.  As the development of hybridization breaks both spin-rotation and time-reversal symmetry, it can only develop at a \emph{phase transition}.  So hybridizing with a non-Kramers doublet leads to a two-channel Kondo problem, converting the normal Kondo crossover into a phase transition.  This idea naturally explains how the hybridization gap can turn on as an order parameter at $T_{HO}$ - it turns on as an order parameter because \emph{it is an order parameter}.  In our microscopic calculation, we find that when the spinor points along the c-axis (and is staggered), the resulting state is AFM with large f-electron moments; however, when the moment points in the basal plane, the resulting state has no large moments and strongly resembles HO.  We have dubbed this state ``hastatic order.''\cite{Chandra13}

We can develop a Landau theory of this spinorial order parameter, $\hat \Psi = (\Psi_\up, \Psi_\down)$:
\begin{equation}\label{}
f[\hat \Psi] = 
\alpha (T_{c}-T)\vert \hat\Psi \vert^{2}+ \beta \vert \hat\Psi \vert^{4} - 
\lambda (\hat\Psi \dg \sigma_{z}\hat\Psi )^{2}.
\end{equation}
The additional term, $\lambda  (\hat\Psi \dg \sigma_{z}\hat\Psi )^{2}$ favors the spinor pointing along the c-axis or within the basal plane, giving the observed HO to AFM transition as a spin-flop of the hybridization spinor from the basal plane to the c-axis.  Moreover, we can calculate the gap to longitudinal spin fluctuations in the hastatic state, and find that it vanishes with a square-root behavior at the first order transition to the AFM. 


\section{Microscopic Theory}

\figwidth=15cm
\fg{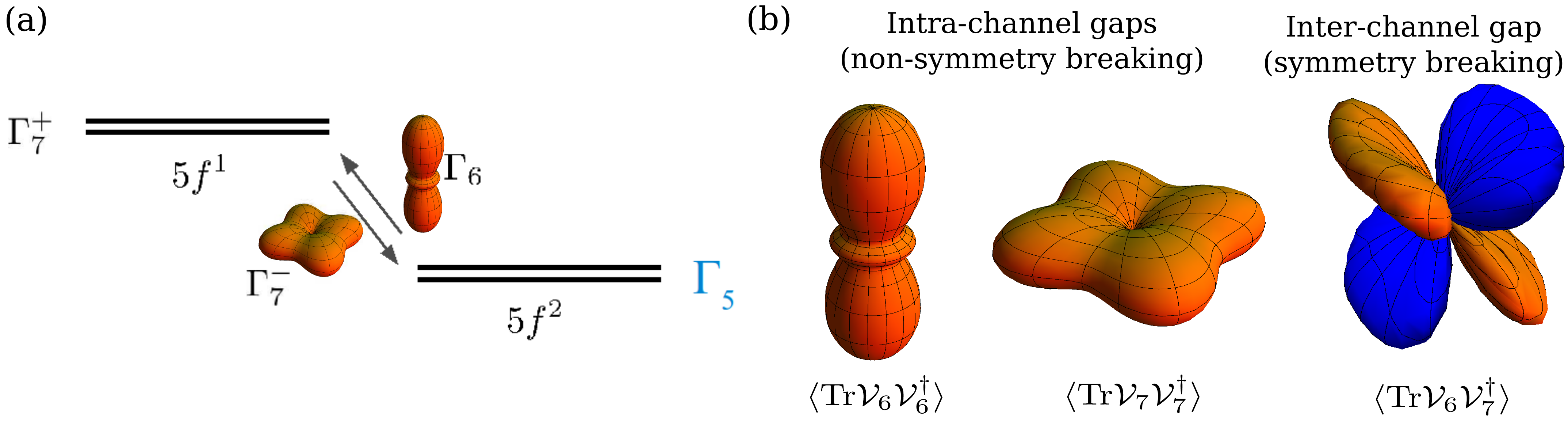}{anderson}{(a) If the ground state of URu$_2$Si$_2$ is the $\Gamma_5$ non-Kramers doublet, then the valence fluctuations will be described by a two-channel Anderson model that is strongly constrained by the form of the doublet.  Here we choose the excited doublet to be the $\Gamma_7^+$ doublet of 5f$^1$, which means that $\Gamma_6$ and $\Gamma_7^-$ conduction electrons will hybridize with the ground state $\Gamma_5$ doublet.  The small figures show the symmetry of these two types of conduction electrons.  (b) There are two types of hybridization gaps resulting from hastatic order: uniform intra-channel gaps that originate with the amplitude of the spinorial hybridization, $\langle \Psi^\dagger \Psi \rangle$ and a staggered inter-channel gap that comes from the phase of the hybridization, $\langle \Psi^\dagger \vec{\sigma}\Psi \rangle$.  Only the last gap breaks time-reversal and tetragonal symmetries.  In the mean field theory, all of these gaps will develop at $T_{HO}$, but beyond mean field theory, only the symmetry breaking gap is required to develop at $T_{HO}$.  We strongly suspect that hastatic order will ``melt'' via phase fluctuations, such that the non-symmetry breaking gaps will survive above $T_{HO}$ and give rise to the hallmarks of heavy fermion behavior.}

The relevant non-Kramers doublet for \urs is the $\Gamma_5$ doublet [see eqn(1)].  To treat the hybridization with conduction electrons, we use a two-channel Anderson model\cite{CoxZawad}, $H =  \sum_{\bk \sigma} \epsilon_\bk c\dg_{\bk \sigma} c_{\bk \sigma} + \sum_j \left[H_a(j) + H_{VF}(j)\right]$, where $H_a(j) = \Delta E \sum_\pm|\Gamma_7 \pm,j\rangle \langle \Gamma_7 \pm,j|$, as $\Gamma_7^+$ is taken to be the lowest energy doublet of the excited 5f$^1$ state (as shown in Fig. 3(a)).  The form of the valence fluctuation Hamiltonian, $H_{VF}(j)$ is completely determined by the structure of the $\Gamma_5$ doublet,
\begin{equation}
H_{VF}(j) = V_6 \psic_{\Gamma_6 \pm}\dg(j) |\Gamma_7^+
\pm\rangle \langle \Gamma_5 \pm| + V_7 \psic_{\Gamma_7 \mp}\dg(j)
|\Gamma_7^+ \mp \rangle\langle \Gamma_5 \pm| + \mathrm{H.c.}.
\end{equation}
The conduction electrons are in $J=5/2$ Wannier states with $\Gamma = \Gamma_6$ or $\Gamma_7^-$ symmetry, $c_{\Gamma \pm}\dg = \sum_\bk \left[\Phi_\Gamma\dg(\bk)\right]_{\pm, \sigma} \mathrm{e}^{-i\bk\cdot \bR_j} c_{\bk \sigma}\dg$.  As we assume the U sites are either in a 5f$^2$ (ground state) or a 5f$^1$ (excited state with energy $\Delta E$), but not a 5f$^3$ state, we must use Hubbard operators to represent the various  configurations.  To make computational progress, we factorize the Hubbard operators using slave bosons, $\Psi_\pm$ and pseudo-fermions, $f_\pm$ to represent the excited Kramers and ground state non-Kramers doublets, respectively, so that $|\Gamma_7^+\alpha,j\rangle \langle \Gamma_5 \beta,j|$ is replaced by $\Psi\dg_{j\alpha} f_{j\beta}$.  We then take the mean-field approximation and condense the slave bosons, $\langle\Psi_{j\alpha}\rangle$, leaving behind a quadratic Hamiltonian describing the hybridization of conduction electrons and pseudofermions.  In the hastatic state, where the hybridization spinor is in-plane and staggered between planes, $\langle \Psi_\pm \rangle = |\Psi| \rm{e}^{\pm i \left(\bQ\cdot \bR_j + \phi\right)/2}$.  The resulting complicated Hamiltonian can be rewritten using the gauge symmetries of the problem as a uniform and staggered hybridization with $\Gamma_6$ and $\Gamma_7^-$ electrons, respectively:
 \bea 
H_{VF} = \sum_\bk c\dg_\bk
\mathcal{V}_6(\bk) f _\bk + c\dg_\bk \mathcal{V}_7(\bk) f _{\bk+\bQ} +
\mathrm{h.c.}  
\eea
In writing the hybridization form factors $\mathcal{V}_7(\bk) = V_7 \Phi_7{\dg}(\bk) \sigma_1$ and $\mathcal{V}_{6}(\bk )= V_{6}\Phi_{6}{\dg} (\bk )$, we have suppressed the spin indices.  If we use $\vec{\tau}$ to represent $\bk$ and $\bk+\bQ$ space, we can compactly write the hybridization matrix as $\mathcal{V}(\bk) = \mathcal{V}_6(\bk) \tau_3+\mathcal{V}_7(\bk) i \tau_2$\cite{note}.  The hybridization matrix itself, $\mathcal{V}$ is not gauge invariant - only the products: $\Tr \mathcal{V}_7 \mathcal{V}_7\dg$, $\Tr \mathcal{V}_6 \mathcal{V}_6\dg$, and $\Tr\mathcal{V}_6 \mathcal{V}_7\dg$ are gauge invariant; these describe observable hybridization gaps and lead to a number of observable consequences.

\subsection{Experimental Consequences}

\subsubsection{Moments}

As hastatic order breaks time-reversal symmetry, one might expect magnetic moments to develop.
In fact, there are three types of moments in the problem: the original $\Gamma_5$ moments, $f\dg \vec{\sigma} f$, which are dipolar along the c-axis and quadrupolar ($\mathcal{O}_{xy}$ and $\mathcal{O}_{x^2-y^2}$) in-plane; conduction electron moments, $c\dg \vec{\sigma} c$; and ``mixed valent'' moments due to the non-zero occupation of the 5f$^1$ state, $\Psi\dg \vec{\sigma} \Psi$.  Below $T_{HO}$, we find that the $\Gamma_5$ moments remain zero, while the conduction electron and mixed valent moments become non-zero and staggered in the basal plane.  As the moment magnitudes originate in the Kondo effect, their small magnitude is dictated by the Kondo energy scale, $O(T_K/D)$\cite{Anderson61}.  In our original calculation\cite{Chandra13}, we took the amount of mixed valency, $\langle \Psi\dg \Psi \rangle$ to be 20$\%$ and predicted a basal plane moment on the order of $.01\mu_B$.  Such a moment has now been ruled out by neutron experiments\cite{Das13,Metoki13}.  However, if the amount of mixed valency is significantly smaller, we can obtain moments on the order of $.0005\mu_B$ that are consistent with those suggested by x-ray scattering\cite{Caciuffo} and the fields seen in NMR\cite{Bernal04}.  If these moments do exist, why the relevant degree of mixed valency is much smaller than the expected from room temperature measurements, is an open question.

\subsubsection{Broken Tetragonal Symmetry}

As there are no f-electron moments, there is no quadrupole moment associated with hastatic order.  However, there is a more subtle tetragonal symmetry breaking due to the mixing of the two hybridization channels, $\Tr\mathcal{V}_6 \mathcal{V}_7\dg$.  This hybridization gap (shown in Fig. 3(b)) clearly breaks tetragonal symmetry, but with a nodal structure that prevents direct coupling to strain and thus quadrupole moments.  However, higher order couplings should lead to a parasitic and small orthorhombic distortion, as seen\cite{Matsuda}, and to a predicted resonant nematicity (centered around the hybridization gap energy $\sim T_{HO}$) in STM.  This tetragonal symmetry breaking also manifests as a nonzero $\chi_{xy}$ in the conduction electrons developing as $(\mathcal{V}\vec{\sigma}\mathcal{V}\dg)^2$.  Since this susceptibility develops as the result of resonant scattering of conduction electrons off the f-electrons, it is relatively large. 

\subsubsection{Hybridization Gap}

One of the key components of hastatic order is the symmetry breaking hybridization gap developing at $T_{HO}$.  The hybridization gap actually has multiple components: the intra-channel gaps, $\Tr \mathcal{V}_7 \mathcal{V}_7\dg$ and $\Tr \mathcal{V}_6 \mathcal{V}_6\dg$ and the inter-channel gap,  $\Tr\mathcal{V}_6 \mathcal{V}_7\dg$.  Only the last gap actually breaks time-reversal and tetragonal symmetries: this gap is associated with the \emph{direction} of the hastatic spinor, $\langle \Psi\dg \vec{\sigma}\Psi \rangle$, while the intra-channel gaps are associated with its \emph{amplitude}, $\langle \Psi\dg \Psi\rangle$.  As the amplitude breaks no symmetries, it can develop above $T_{HO}$ as a crossover (beyond mean-field theory), just as in the usual Kondo effect, and so it will lead to the development of heavy fermion physics above $T_{HO}$, as seen in \ursp.  However, the inter-channel gap must develop at $T_{HO}$, as seen in STM.  This inter-channel gap has a roughly d-wave character oriented along the $xy$ axis, with nodes in the $k_z = 0$ plane and the $k_y = -k_z$ planes, which should be observable in ARPES and other band-structure measurements.

\figwidth=8.5cm

\section{Conclusions}

The essence of our hastatic order proposal for \urs is that dual observations of Ising quasiparticles and identification of the hybridization and hidden order gaps in STM naturally require the hybridization of a non-Kramers doublet with Kramers conduction electrons.  As this hybridization mixes integer and half-integer spin states, it must itself carry a half-integer spin, and thus behave like a spinor under time-reversal.  This sort of spinorial hybridization naturally explains the large entropy of condensation (originating from the Kondo effect); absence of large moments; broken tetragonal symmetry and Ising quasiparticles.  Hastatic order also predicts a vanishing gap to longitidinal spin fluctuations at the HO/AFM transition; tiny staggered transverse magnetic moments in the basal plane; and a nodal, d-wave hybridization gap develop at $T_{HO}$, resulting in a resonant nematicity.


\begin{thebibliography}{9}

\bibitem{Palstra85} T.T.M. Palstra et al., {\bf Phys. Rev. Lett.} {\bf 55} 2727 (1985).

\bibitem{Schlabitz86}W. Schlabitz et al., {\bf Z. Phys. B. 62}, 171 (1986).

\bibitem{Mydosh11}J. A. Mydosh and P. M. Oppeneer, {\bf Rev. Mod. Phys. 83}, 1301 (2011).

\bibitem{Chandra13} Premala Chandra, Piers Coleman and Rebecca Flint, {\bf Nature 493}, 621 (2013).

\bibitem{sakurai}J. J. Sakurai, ``Modern Quantum Mechanics'', Revised Edition, pp 277, (Addison-Wesley), (1994).

\bibitem{CoxZawad}D. L. Cox and A. Zawadowski, ``Exotic Kondo Effects in Metals'', Taylor \& Francis, London. (2002).

\bibitem{Amitsuka94} H. Amitsuka and T. Sakakibara, {\bf J. Phys. Soc. Japan} {\bf 63} 736 (1994).

\bibitem{PiersReview} P. Coleman, ``Heavy Fermions: Electrons at the Edge of Magnetism,'' Handbook of Magnetism and Advanced Magnetic Materials (2007).

\bibitem{Coleman83} P. Coleman, 
{\bf Phys. Rev. B 29}, 3035 (1984).

\bibitem{Read83}N. Read, and D. M. Newns, 
{\bf J. Phys. C} {\bf 16}, L1055 (1983); \emph{ibid} 3273 (1983).

\bibitem{Jo07}Y.J. Jo et al.,
{\bf Phys. Rev. Lett.} {\bf 98}, 166404 (2007).

\bibitem{Villaume08} A. Villaume et al., 
{\bf Phys. Rev. B} {\bf 78} 
5114 (2008).

\bibitem{Hassinger10} E. Hassinger et al, 
{\bf Phys. Rev. Lett.} {\bf 105}, 216409 (2010).

\bibitem{Broholm91} C. Broholm et al., 
{\bf Phys. Rev B.} {\bf 43}, 12809 (1991).

\bibitem{Niklowitz11}P. G. Niklowitz et al., 
arXiv:1110.5599 (2011).

\bibitem{Okazaki11} R. Okazaki et al.,  {\bf Science} {\bf 331} 439 (2011).

\bibitem{Matsuda} Private communication from Y. Matsuda and T. Shibauchi.

\bibitem{Schmidt10} A. R. Schmidt et al,  {\bf Nature} {\bf 465}, 570 (2010).

\bibitem{Aynajian10}P. Aynajian et al., {\bf PNAS} {\bf 107}, 10383 (2010).

\bibitem{lanl} G. L. Dakovski et al, {\bf Phys. Rev. B 84}, 161103(R) (2011).

\bibitem{Shen13}  S. Chatterjee et al, {\bf Phys. Rev. Lett. 110}, 186401 (2013).

\bibitem{Ohkuni99} H. Okhuni et al.,
Philos. Mag. B 79, 1045 (1999).

\bibitem{Altarawneh11} M.M. Altarawneh et al., 
{\bf Phys. Rev. Let.} {\bf 106}, 146403 (2011).

\bibitem{Hc2}J. P.  Brison et al, 
 {\bf Physica C 250}, 128 (1995).

\bibitem{Altarawneh2}M. M. Altarawneh et al. , 
 {\bf Phys. Rev. Lett. 108}, 066407 (2012). 

\bibitem{Park02} J.-G. Park, K. A. McEwen, M. J. Bull, {\bf Phys. Rev. B} {\bf 66}, 094502 (2002).

\bibitem{Jeffries10} J.R. Jeffries et al., {\bf Phys. Rev. B} {\bf 82}, 033103 (2010).

\bibitem{Flint13} Rebecca Flint, Premala Chandra and Piers Coleman, {\bf Phys. Rev. B 86}, 155155 (2012).

\bibitem{note} This compact notation is possible since the form-factors obey $\Phi_\Gamma\dg(\bk + \bQ) = -\Phi_\Gamma\dg(\bk)$ for commensurate $\bQ$.

\bibitem{Anderson61}P.W. Anderson, 
{\bf Phys. Rev. 124}, 41 (1961). 

\bibitem{Das13} P Das et al, {\bf New J. Phys. 15}, 053031 (2013).

\bibitem{Metoki13} N. Metoki et al, {\bf J. Phys. Soc. Jpn. 82},055004  (2013).

\bibitem{Caciuffo} Private communication from R. Caciuffo.

\bibitem{Bernal04} O.O. Bernal et al., 
{\bf J. Mag. Magn. Mat. 272}, {E59-60} (2004).



\end{thebibliography}
\end{document}